\documentclass{article}
\usepackage{graphicx}  
\usepackage{amsmath}   
\usepackage[compress]{cite}
\usepackage{amssymb}   
\usepackage{bm} 
\usepackage{dcolumn}
\usepackage{color}
\usepackage{mathrsfs}
\usepackage{amsfonts}
\usepackage{varioref}
\RequirePackage[colorlinks,citecolor=blue,urlcolor=magenta,linkcolor=blue]{hyperref}
\newcommand{\be}{\begin{equation}}
\newcommand{\ee}{\end{equation}}
\newcommand{\bea}{\begin{eqnarray}}
\newcommand{\eea}{\end{eqnarray}}

\def\EH{Einstein-Hilbert }
\def\LL{Lanczos-Lovelock }
\def\gr{general relativity}
\labelformat{section}{Section #1} 
\labelformat{subsection}{Section #1} 
\labelformat{subsubsection}{Section #1}
\labelformat{subsubsubsection}{Section #1}
\labelformat{equation}{Eq.~(#1)} 
\labelformat{figure}{Fig.~#1} 
\labelformat{subfigure}{Fig.~\thefigure#1} 
\labelformat{table}{Tab.~#1} 
\labelformat{appendix}{Appendix #1}
\title{Discrete quantum spectrum of black holes}
\author{Kinjalk Lochan
\footnote{kinjalk@iucaa.in}
\hskip 2mm
and
\hskip 2mm
Sumanta Chakraborty
\footnote{sumanta@iucaa.in; ~~sumantac.physics@gmail.com}\\
{\small{IUCAA, Post Bag 4, Ganeshkhind,}}\\
{\small{\it Pune University Campus, Pune 411 007, India}}}
\begin{document}
  
\maketitle
\begin{abstract}
The quantum genesis of Hawking radiation is a long-standing puzzle in black hole physics. Semi-classically one can argue that the spectrum of radiation emitted by a black hole look very much sparse unlike what is expected from a thermal object. It was demonstrated through a simple quantum model that a quantum black hole will retain a discrete profile, at least in the weak energy regime. However, it was suggested that this discreteness might be an artifact of the simplicity of eigen-spectrum of the model considered. Different quantum theories can, in principle, give rise to different complicated spectra and make the radiation from black hole dense enough in transition lines, to make them look continuous in profile. We show that such a hope from a geometry-quantized black hole is not realized as long as large enough black holes are dubbed with a classical mass area relation in any gravity theory ranging from GR, Lanczos-Lovelock to f(R) gravity. We show that the smallest frequency of emission from 
black hole in any quantum description, is bounded from below, to be of the order of its inverse mass. That leaves the emission with only two possibilities. It can either be non-thermal, or it can be thermal only with the temperature being much larger than 1/M. 
\end{abstract}
\section{Introduction}

General relativity is a very successful theory and so far is the best candidate to describe the geometrical properties of the spacetime. The success of general relativity crucially hinges on the fact that it has passed through all the experimental and observational tests so far. The observational tests span a wide range of parameter space, starting from local gravity tests like perihelion precession and bending of light to high precision tests using pulsars. Despite of these outstanding successes for \gr\ there are quiet a few unresolved issues, e.g., the problem of dark energy and the problem of inflation. These prompted research in a new direction by modifying the Einstein-Hilbert action for \gr\ itself. Among various alternatives the criteria that the field equations should remain second order in the dynamical variable (otherwise some ghost fields would appear) uniquely fixes the action to be the Lanczos-Lovelock action \cite{gravitation,Padmanabhan:2013xyr}. Another way to arrive at the same is to 
generalize the curvature tensor such that trace of its Bianchi derivative vanishes, which yields a divergence free second rank tensor uniquely leading to the \LL Lagrangian \cite{Dadhich:2008df}. The pure \LL Lagrangian, i.e., one particular order out of the full \LL Lagrangian is closely associated with spacetime dimensions as well. For example, field equations for \gr\ are non-trivial for $D>2$ while it has free propagation only in $D>3$. For $D=3$, a peculiar phenomenon happen, Riemann tensor gets determined entirely by Ricci tensor and gravity becomes kinematic. If we insist that the kinematic property of gravity should hold in all odd dimensions, then it uniquely singles out pure \LL gravity \cite{Dadhich:2015lra,Yale:2010jy}. What is more, from thermodynamic perspectives as well \LL gravity has a special status, since most of the thermodynamic results holding in \gr\ can be generalized to \LL gravity as well \cite{Padmanabhan:2013nxa,Chakraborty:2014rga,Chakraborty:2014joa,Chakraborty:2015wma,
Chakraborty:2015hna}. 

Another such model with the potential of explaining the above mentioned problems can be achieved by replacing $R$, the scalar curvature in the \EH action by some arbitrary function of the scalar curvature $f(R)$. This alternative theory has the potential to pass through local gravity tests and can explain a variety of phenomenon including the late time cosmic acceleration \cite{Nojiri:2010wj,Sotiriou:2008rp,DeFelice:2010aj}. Alike, $f(R)$ theory, we can invoke ``teleparallelism'' and construct a $f(T)$ theory of gravity, where $T$ stands for torsion scalar which can be regarded as another alternative to \gr. Also in cosmological scenarios this model allow for both inflation and late time cosmic acceleration. Black holes and their entropy is also another well studied subject in the context of $f(T)$ gravity where the Bekenstein area law is obtained as a leading order correction \cite{Chakraborty:2012kj,Harko:2011kv,Sharif:2014kda,Wang:2011xf,Iorio:2012cm}.  

In all these alternative gravity theories, the most fascinating objects appearing as solutions of the respective gravitational field equations are the black holes. Despite the fact that black holes are supposed to be perfect trapping systems at the classical level, there are enough evidences to point out that they seem to behave much like a thermal object \cite{Bekenstein:1972tm,Unruh:1983ms}. This belief was strengthened once Hawking showed that black holes seem to be radiating at a characteristic temperature inversely proportional to their masses \cite{Hawking:1974sw}. Therefore, similar to a thermal body, a black hole is also prescribed to have some entropy which turns out to be proportional to their area of horizon in the Einstein theory or proportional to some powers of area in \LL theories of gravity \cite{Dadhich:2012ma,Paranjape:2006ca}. In the case of \gr\ one can look for this intriguing area dependence via the entanglement entropy for the black hole \cite{Eisert:2008ur,Brustein:2005vx,Calabrese:
2005zw,Masanes:2009tg}. As and when a correct theory of quantum gravity is achieved, a natural onus on it will be to show the corresponding microstates in a black hole making up for the entropy. Not only this, a consistent quantum theory of gravity is also expected to shed light on the thermal behavior of black holes from microscopic point of view.

All these results follow from semi-classical arguments and are studied extensively for observers in the regions outside black hole event horizon. Recently in \cite{Chakraborty:2015nwa,Singh:2014paa} it has been shown that energy density of a quantum field dominates over the classical background, near the singularity and have the potential to alter the singularity structure. Also in \cite{Lochan:2015oba} the semi-classical approximation is shown to receive correction leading to some non-trivial implications for information content in the Hawking radiation (see also \cite{Page:1993wv,Hawking:1976ra,Mathur:2009hf,Mathur:2008kg}). All these results motivate us to study the semi-classical thermal Hawking spectrum in a full quantum picture, if possible. One way to do that will be to treat the black hole as a highly excited quantum object\footnote{In fact there are prescriptions like {\it fuzzballs} \cite{Mathur:2008kg} that suggest there should not be any consistent fully classical description of black holes and 
they are indeed purely quantum mechanical objects.} or a macroscopic construct of a consistent microscopic theory of gravity. A large black hole can then be thought of as either a highly excited state of the fundamental quantum description or a macroscopic limit of the underlying quantum description (i.e. $N\rightarrow \infty$ where $N$ is the total number of quanta constituting the hole).

Although there are semi-classical strong arguments suggesting a thermal behavior of Hawking radiation, there could be various factors which can distort the thermal profile of the radiation. There are various classical phenomenon, e.g., grey-body factor, non-adiabaticity which can produce distortions in the Infra-red regime of the Hawking spectra \cite{Visser:2014ypa}. Recently, there has been a renewed interest \cite{Gray:2015pma,Hod:2015wva,VanPutten:2015pja} in early-stage low-temperature regime of black hole evaporation. More particularly, it has been argued previously within the context of \gr\ \cite{Gray:2015pma,Hod:2015wva} that macroscopic black holes should have large elapse time resulting in a sparse spectrum. One may expect from a potential quantum theory of black holes, that some of these properties might find some backing from quantum point of view. One can also expect that the essential classical characteristic features of black holes get quantized too, in a quantum theory of gravity. It has 
been argued by Bekenstein et. al. \cite{Bekenstein:1995ju} that if one takes the black hole as a highly excited state of the quantum description, the 
emission profile of the Hawking radiation remains very sparse such that in the dominant part of the thermal spectra only a few lines contribute and most of the continuum feature is present towards the tail. So {\it the Hawking radiation is quantum mechanically silent in the region where bulk of the thermal radiation comes from.} This effect was taken as a stumbling block for some earlier quantization models \cite{Bekenstein:1973ur,Hod:2000it,Hod:1998vk} which predicted the area of the hole to be quantized in integer steps. However, it was still to be seen whether such a  phenomenon is general enough to accommodate different quantum spectra arising from different quantum gravity theories.

In this letter, we argue that the sparseness of Hawking radiation survives the modification of quantum eigen-spectrum of geometric observables for a large class of gravity theories, which includes \gr. So, the sparseness of Hawking spectra has nothing to do with precise spectral description of the underlying quantum theory\footnote{It was argued \cite{Rovelli:1994ge} in the framework of Loop Quantum Gravity (henceforth referred to as LQG), that the area spectrum remains no more equidistant and gets quantized like an angular momentum operator. This supposedly filled in the sparse region in the radiation spectrum in the equidistant case, with additional states, introducing more number of transition lines thus populating the discrete profile, making it look more continuous.}. We show that the following (semi)-classical inputs --- (1) {\it area entropy relation} determined by the gravity model we are interested in (in particular for \gr\ the entropy of the hole is directly proportional to its area), (2) {\it 
classical mass area} relation and (3) {\it an effectively continuous Hawking spectra}--- are incompatible with each other in a full 
quantum theory. We have explicitly demonstrated that quantum spectrum of black holes for arbitrary quantization schemes in \gr, Lanczos-Lovelock, $f(R)$ and $f(T)$ theories are all bounded below by inverse the black hole mass. Thus the inverse mass cutoff associated with black hole quantum spectrum is a generic result, since it holds for a very general quantization scheme and for a large class of gravity theories.

The letter is organized as follows, in \ref{Quant_GR} we will first argue for the general case within the \gr\ premise where a black hole gets quantized by some underlying quantum theory (which we are not interested in, presently). Thereafter, we will demonstrate the situation for \LL theories of gravity along with $f(R)$ and $f(T)$ theories in \ref{Quant_LL}. Finally we conclude with a discussion. Also demonstration of some specific quantization scheme for some particular geometric quantities of the black hole have been performed in \ref{App_Quant}. 
\section{Quantized Black hole}\label{Quant_GR}

Let us start by describing a black hole as a macroscopic system, such that the quantization of some geometrical observable $\mathcal{Z}$ of black hole in the characteristic theory (for instance such as in \cite{Vaz:2007td,Vaz:2009jj}), corresponds to the following,
\bea
{\cal Z}= \alpha \sum_jn_j f(j),\label{Arb_Spec}
\eea
with an arbitrary $f(j)$. This is a reminiscent of a statistical behavior of the system where the expectation value of the operator $\hat{\cal Z}$ is used as a variable. We will consider the case where in the macroscopic limit the area of the event horizon gets related to this characterizing geometric parameter as (for some explicit examples see \ref{App_Quant})
\bea
A \propto {\cal Z}^{\gamma} \Rightarrow A = \alpha ^{\gamma} \left(\sum_jn_j f(j)\right)^{\gamma}. \label{A-Z Rel}
\eea
Again quantum mechanically, the area operator would have been dependent on $\hat{\cal Z}^{\gamma}$ and hence the expectation value would have been 
a different functional of spectral profile $f(j)$, but the analysis goes through exactly as described below, without altering the outcome. Secondly, the difference between these two functionals will vanish in large $N$ approximation. We will show that even with this most general choice we would be able to get the lowest frequency to be $\sim 1/M$. 

For large enough black hole, the characterizing geometric variable should be determined in terms of its mass (through a mass-area relation), in order to respect the {\it no-hair} conjecture. Further, the Bekenstein-Mukhanov case can be obtained from \ref{A-Z Rel} by choosing $\gamma=1$, $n_{j}=\delta_{ij}$ and $f(j)=j$. If we assume in the spirit of \cite{Bekenstein:1995ju}, that there is a {\it holographic} description of black holes, i.e., its entropy and area are related by,
\bea
S &=& \frac{A}{4} \label {Holography}\\
  &=& \frac{\alpha^{\gamma}\left(\sum_jn_j f(j)\right)^{\gamma} }{4} = \log{g(n)},
\eea
where $g(n)$ is effective number of microstates giving rise to the black hole configuration. This effective number comes from thinking the black hole either as a microscopic limiting case of the quantum description or the degeneracy of the higher excited state if that be the case. From \ref{Holography} it is simple to compute the number of microstates, which leads to,
\bea
g(n)=\exp{S}=\exp\left({\frac{A}{4}}\right)=\exp\left[{\frac{\alpha^{\gamma}}{4} F^{\gamma}}\right];\qquad F=\left(\sum_jn_j f(j)\right).\label{IntegerConstraint}
\eea
Clearly, for the right hand of side of \ref{IntegerConstraint} to be able to churn out an integer, we require $\alpha^{\gamma}/4 =\log{K}$ for some integer $K$. But additionally, we have another constraint that $F^{\gamma}$  has to be an integer for all possible $\{n_j\}$ (at least) for which $\sum_j n_j \gg 1$. Otherwise, the integer criterion, i.e., \ref{IntegerConstraint} will not work out for arbitrary high occupancies. We consider a Schwarzschild like massive black hole admitting the classical area-mass relation\footnote{In fact in LQG, this logic is adopted to argue for new mass states by LQG scheme of quantization \cite{Barreira:1996dt}, where the presumed number of degenerate area states are argued to be exponentially large. However, in general, a quantization scheme can give rise to multiple number of degenerate states, it will not shift the mass separation gap if such a mass-area relation is adopted. We show that modification to this relation offers very little help for filling the emission profile 
of the black hole. Moreover the Hardy Ramanujan relation for the degeneracy of mass states is not applicable once the parts are assumed to carry large spin values.}
\bea
M_n &\propto& A_n^{1/2} \label{M-A Rel} \\ 
&\propto& \alpha^{\gamma/2} F^{\gamma/2}. 
\eea
The emission profile of the black holes will be obtained from the transitions to lower mass eigenstates which will be obtained as the quasi-normal frequency \cite{Hod:2000it,Hod:1998vk} associated with the black hole geometry. From \ref{M-A Rel}, we have
\bea
\frac{\Delta M}{\hbar}  \propto \frac{F^{\gamma/2}- F'^{\gamma/2}}{\hbar}.
\eea
Here $F$ stands for occupancy $\{n_{j}\}$ and $F'$ stands for $\{n_{j'}\}$. From holography we know that the quantity $F^{\gamma}$ has to be integer for all possible $\{n_j\}$ which satisfy $\sum_j n_j \gg 1$ for a stable system. Thus mass difference turns out to be difference of  square roots of two such integers constructed from two different set of occupancies $\{n_j\}$ and  $\{n_j'\}$. Clearly, the smallest, the integers could change would be by unity and hence the smallest frequency of radiation for a massive black hole will be
\bea
\omega = \frac{\Delta M}{\hbar} \propto \frac{I^{1/2}- \left(I-1\right)^{1/2}}{\hbar} \propto \frac{1}{\hbar I^{1/2}}\sim \frac{1}{M},
\eea
where we have written $F^{\gamma}=I$. Therefore, the leading order frequency dependence of the emitted radiation goes as $\omega \propto 1/M$. We note here, the precise spectral form of the geometric variable  was never required for this demonstration. The criterion $I^{\gamma}$ being an integer will not be satisfied for all arbitrary $f(j)$ and $\gamma$, but again this criterion is mandatory for a quantum gravity theory which will explain the holographic character of the event horizon of the black hole. It can be an interesting number theory exercise to obtain possible pairs of $(f(j),\gamma)$ leading to a consistent holographic description. Secondly, it is not clear which kind of geometric spectra will provide a thermal envelope to the emission profile. We will not comment further on these issues and pursue them elsewhere. However, for the moment, adopting such a spectra and the constraint, rest of the arguments follow and with the classical mass area relation we again obtain the same dependence of 
smallest quasi-normal modes being $1/M$. So we show that irrespective of details of quantization, the sparseness of Hawking spectrum emerges as a robust feature which can be applied to any geometric quantization scheme and which yields a classical {\it holographic} and mass area relations. We will see that it is necessary to give up one of these two characterizing features of the macroscopic black hole, if denseness of radiation is to be obtained.

As pointed out in \cite{Padmanabhan:Private,Bekenstein:1974jk,Kotwal:2002ch} the mass dependence of the minimum frequency quanta radiated by the black hole can also be related to the ``cavity size''\footnote{We note that even in standard black body radiation there will be an apparent cutoff when the wavelength of the black body radiation $\sim$ the size of the cavity.} of the black hole. However in the case of any covariant gravity theory the cavity size of a black hole can depend on the slicing of the spacetime manifold (in particular the region interior to the black hole) \cite{DiNunno:2008id}. Even though the cavity size may differ depending on the slicing, but to an exterior observer receiving the Hawking radiation the area radius plays the pivotal role, showing the holographic character of black holes.

Given the outcome outlined above, it is pertinent to ask, what are the possible ways in which we could have generated a dense spectrum. There two possibilities, which we will discuss now. 
\begin{itemize}
 
\item Firstly, by altering the classical mass-area relation it is possible to generate a dense spectrum. The best way to see this is to start with a mass-area relation of the form $M_{n}=A^{\chi /2}$, where $\chi =1$ yields the standard classical mass-area relation. In this case, by straightforward algebra it turns out that lowest frequency corresponds to $\omega =M^{1-(2/\chi)}$. Thus for $\chi =1$ we have the $\omega \sim (1/M)$ result as obtained before. However if the mass-area relation were different and $\chi<1$, then it would be possible to get a dense Hawking spectrum. 

\item Another possibility to obtain a dense enough spectrum will be to modify the holographic area entropy relation. This can be achieved if the theory of gravity which is chosen for quantization is not Einstein's General Relativity. In that case such an area entropy relation can also naturally change. It will be relevant to study different modified gravity theories to see the denseness of emitted radiation from black hole solutions therein. This is what we will study in the next section, which includes \LL gravity, $f(\mathcal{R})$ gravity, $f(T)$ gravity etc. It will be shown that despite all these efforts the $1/M$ cutoff is always existing.  
 
\end{itemize}
\section{Generalization to Lanczos-Lovelock Gravity}\label{Quant_LL}

So far we have been within the premise of general relativity and we found that the quantum spectrum of black hole is discrete with $1/M$ being the low frequency cutoff, which coincides with the peak of the Hawking spectrum. A natural question emerges out of this discussion, what happens if the entropy no longer scales as area. It is well known that entropy of black holes depend crucially on the gravity theory we are interested in. Among all the other alternative theories the \LL gravity is of quiet importance, due to its similarity with \gr\ (as this also yields second order field equations) and its thermodynamic special status. Given this we will elaborate on the \LL gravity and finally shall provide some discussion of discrete Hawking spectrum regarding $f(R)$ and $f(T)$ theories of gravity. 

It is well known that entropy of black holes in \LL gravity does not scale with area \cite{Dadhich:2012ma,Majhi:2011ws}. Given this it is not clear a priori whether the resulting Hawking spectra would be dense or not. To investigate this question we will stick to the most general quantization scheme for a black hole, i.e., the one presented in \ref{Arb_Spec}. In general \LL gravity can be written as a sum of various terms, such that $m$th term is a homogeneous function of the Riemann tensor of order $m$. We will start with pure Lovelock black holes of order $m$ (which corresponds to a single term in the full series), in $d$ spacetime dimensions. For the pure Lovelock black holes, the black hole horizon, in the simplest situation is located at $r=r_{h}$, is related to the ADM mass by the relation \cite{Kastor:2008xb},
\begin{align}
r_{h}=M^{1/(d-2m-1)}.
\end{align}
For $d$ dimension, the black hole entropy for pure Lovelock black hole can be calculated and it corresponds to \cite{Dadhich:2012ma,Paranjape:2006ca},
\begin{align}
S=r_{h}^{d-2m}.
\end{align}
Let some geometrical variable $\mathcal{Z}$ gets quantized according to the quantization rule,
\begin{align}
\mathcal{Z}=\alpha \sum _{j}n_{j}f(j);\qquad A=\mathcal{Z}^{\gamma}.
\end{align}
Now area for the black hole in d-dimension scales as, $A=r_{h}^{d-2}$. Hence the entropy can be written as,
\begin{align}\label{LL_Main}
S=\log g(n)=A^{(d-2m)/(d-2)}=\alpha ^{\gamma(d-2m)/(d-2)}\left(\sum_{j}n_{j}f(j)\right)^{\gamma(d-2m)/(d-2)},
\end{align}
where $g(n)$ stands for multiplicity of states and it has to be integral. Using this criterion we readily obtain, the condition that,
\begin{align}\label{Rev_01}
\alpha ^{\gamma(d-2m)/(d-2)}=\log K,
\end{align}
where $K$ is integral. Using which from \ref{LL_Main} we readily obtain, 
\begin{align}
\left(\sum _{j}n_{j}f(j)\right)^{\gamma(d-2m)/(d-2)}=I,
\end{align}
where $I$ belongs to the set of integers. Then we readily obtain, 
\begin{align}
M=A^{(d-2m-1)/(d-2)}\propto I^{(d-2m-1)/(d-2m)}.
\end{align}
Such that, the lowest quasi normal mode frequency turns out to be,
\begin{align}
\omega \propto \left(I+1\right)^{(d-2m-1)/(d-2m)}-I^{(d-2m-1)/(d-2m)}=I^{-1/(d-2m)}\sim \frac{1}{M^{1/(d-2m-1)}}.
\end{align}
As a quick check, note that for \EH action $m=1$ and thus in four dimension ($d=4$), the quasi-normal mode frequency will scale as $1/M$ as we have shown earlier. However for \gr\ in higher dimension the frequency scales as $1/M^{1/(d-3)}$, resulting in more and more sparseness in the Hawking radiation. Thus \gr\ in four dimension yields the most dense Hawking spectrum.

To our surprise the same is true for \LL gravity as well. For pure Lovelock black holes in even critical $d=2m+2$ dimensions the lowest quasi-normal mode scales as $1/M$, while for all $d>2m+2$, the Hawking spectra more and more sparse. Thus even for pure Lovelock, critical dimension yields the most dense Hawking spectra. 

Let us now turn our attention to Einstein-Lovelock gravity theories. For simplicity we will consider Einstein-Gauss-Bonnet gravity in five dimensions. Then the mass radius relation gets modified and we have $M\sim r_{h}^{2}$. The entropy area relation also gets modified and we obtain, 
\begin{align}
S\sim\frac{r_{h}^{3}}{4}+\frac{3}{2}\alpha r_{h}=\frac{A}{4}+\frac{3 \alpha}{2} A^{1/3},
\end{align}
where $\alpha$ is the Gauss-Bonnet coupling term. When we use the same quantization scheme (see \ref{Arb_Spec}) for some geometric variable $\mathcal{Z}$, related to area by $A=\mathcal{Z}^{\gamma}$ and then try to solve for the area. Since multiplicity has to yield an integer, for large $A$ (i.e., for macroscopic black hole), we will obtain, $\sum _{j} n_{j}f(j)=I^{1/\gamma}$, for integral $I$. Now from the mass radius relation we obtain, $M\sim I^{2/3}$. Thus the lowest quasi-normal mode frequency would scale as, $\omega \sim I^{-1/3}=M^{-1/2}$. Hence for Einstein-Gauss-Bonnet gravity the hawking spectrum is more sparse compared to Einstein gravity itself. This will hold true for all the terms in the Lovelock series. 

This again shows an instant when pure Lovelock gravity in critical dimensions play an important role \cite{Dadhich:2012ma}. In this case we have explicitly shown that most dense Hawking radiation can be obtained from pure Lovelock black hole in the critical spacetime dimensions. However the low frequency cutoff in the Hawking spectrum is still at $1/M$. Hence \LL gravity cannot lead to dense Hawking spectrum. 

To probe the root of this behavior let us start with the following entropy area relation
\begin{align}
S\sim A^{p}=\alpha ^{p\gamma}\left(\sum _{j}n_{j}f(j)\right)^{p\gamma}=\log g(n).
\end{align}
Since the multiplicity of the states $g(n)$ has to be integral by choosing $\alpha$ appropriately as in \ref{Rev_01} we readily obtain, 
\begin{align}
\left(\sum _{j}n_{j}f(j)\right)^{p\gamma}=I,
\end{align}
where $I$ stands for an integer. Then we assume a mass radius relation such that $M\sim A^{\chi}$, hence we obtain, 
\begin{align}
M\sim I^{\chi /p};\qquad \omega \sim I^{\frac{\chi}{p}-1}=M^{(\chi-p)/\chi}.
\end{align}
Hence for dense spectrum we must have $(\chi-p)/\chi<-1$, which eventually leads to, $p>2\chi$. For four dimension $\chi =1/2$, and thus we must have $p>1$. However in all the alternative theories, e.g., $f(R)$, $f(T)$ theories the leading order contribution as always area, all the corrections are sub-leading and hence in none of these gravity theories $p>1$ condition is ever met. Thus none of these theories can lead to a dense spectrum.
\section{Conclusions}

In this paper we try to obtain quantum support for a dense emission profile for a black hole. There have been recent suggestions \cite{Gray:2015pma,Hod:2015wva} for time domain discreteness of Hawking spectrum, while there has been quantum support \cite{Bekenstein:1973ur,Bekenstein:1995ju,Hod:2000it,Hod:1998vk,Hod:2000kb} for a discrete emission spectra of a black hole in the frequency space as well. There is a general hope that including the quantum description of gravity will make black hole a quantum object. So large black holes may support many new transition states which it can jump into and emit a spectral line. Thus, in principle there can be many transition lines and that can make the spectrum rich. However, semi classically, quasi normal modes tell us about the emission profile of the black hole. The frequency of the quasi normal modes can be thought to be the energy extracted by the quantum field in the background spacetime from the black hole. So quantum mechanically the black hole is expected to 
settle down in a new mass state allowed by the quantum theory after emitting one quanta of radiation. So a quantum theory which gives rise to a dense Hawking radiation 
profile should have many nearby mass states for black holes, so that the emission frequency can be extremely small. In other words, the quantum cut-off on frequency in IR regime must be extremely small, so that the empty region in the IR domain as prescribed in semi-classical treatment gets densely occupied.

We consider an arbitrary quantum description of a black hole where one of its characterizing variables gets quantized. We assume that this characterizing parameter at the macroscopic limit should get related to the mass of the hole, owing to the celebrated {\it no-hair} conjecture. For that purpose we take two inputs about a large mass black hole which a classical black hole is expected to satisfy. First we assume that the holography of the black hole is recovered for large $N$ (or large mass) limit. That is to say for such black holes, the entropy, at the leading order, can be given as the macroscopic average area. Secondly, we assume that the mass-area relation for the classical hole is also obtained in that limit. We show that these inputs are sufficient to rule out any quantum geometric description of a black hole if a dense Hawking spectrum is taken as a guiding criterion. Alternatively, one can say that the semi-classical sparseness of a black hole radiation is also supported by the quantum geometric 
approach. This suggests that either the emission profile of the black hole is non-thermal, or, it may only be thermal with a temperature much larger than the IR cut-off which is $1/M$. We have demonstrated this feature not only for \gr\ but also for \LL models of gravity. It turns out that the best one can get is a $1/M$ cut-off and this appears for pure Lovelock theories in critical dimensions. Further we have shown that in four dimensions in order to have dense spectrum the entropy should scale as $A^{n}$, with $n>1$. None of the theories discussed here along with $f(R)$, $f(T)$ theories can have such a behavior and hence the $1/M$ cutoff will appear as a general feature while discussing Hawking radiation in all these alternative theories. Therefore, we argue that a fully quantum mechanical description is unlikely to ascribe the hole with a thermal character with the temperature decided by inverse of its mass. 

It is interesting to ask which kind of geometric quantization can resemble a thermal profile for large mass black holes. We also  propose a constraint on models of geometric quantization which is obtained from the requirement of holographic description of black hole. These two criteria may allow only a few possibilities for underlying quantum spectrum. Furthermore, with a specified area mass relation, area entropy relation can be tweaked to obtain a dense spectrum. However we have explicitly shown that such a relation may be not realized in standard alternative gravity theories, e.g., Lovelock, $f(R)$ and $f(T)$ theories.

Therefore, this analysis suggests that there is an impending non-thermal character to the hawking radiation which will have interesting implication for the infamous information paradox corresponding to the spectral profile. Also, such a behavior can possibly lead to some interesting emission modes of black holes in different gravity theories. Analysis of such issues will be reported elsewhere.
\section*{Acknowledgements}

Research of S.C. is funded by a SPM fellowship from CSIR, Government of India. The authors thank Naresh Dadhich for suggestions on this draft and T. Padmanabhan for illuminating discussions. 
\appendix
\labelformat{section}{Appendix #1} 
\section{Demonstration with simple quantization models}\label{App_Quant}

In the main text we have outlined, how a general characteristic geometric observable of black hole getting quantized, dose not cure the supposed sparseness of the observed Hawking spectrum. Therefore, unlike \cite{Yoon:2012cq}, we do not require a particular observable, like area, to be quantized. We also allow for transitions to all possible quantum states. Our demonstration, remains valid for all quantized geometric variables with all permissible transitions. It is helpful to elucidate the discussion with three illustrative quantization schemes. Since, classically, owing to {\it no-hair} conjecture it is expected that the geometry of a black hole can be characterized by three observable charge, namely mass, electric charge and angular momentum, the black holes carrying one or many of these charges will modify the assumed relation between different macroscopic parameters. However, all of them will still satisfy the holographic relation \ref{Holography} in the Einstein's gravity. Therefore, in the absence of 
a consistent quantum gravity theory describing a black hole fully in quantum mechanical treatment, we study three models of black hole, where different geometric parameters associated with the hole are quantized. As demonstrated, modulo details, the spectral profile separation will 
expectedly tally with the results obtained for the general case in the main text.
\subsection{Mass Quantization}

This model assumes the mass (which in the macroscopic case is marker of the radius for spherical symmetric hole) gets quantized.
Let us start with assuming the black hole to be a quantum system and its mass in the $n-$th quantum level is given by,
\begin{align}
M_{n}=\alpha n^{\beta}.
\end{align}
Then if there is a transition between $(n+1)$ and $n$th level then the mass change is given by (with $n\gg 1$),
\begin{align}
\Delta M&=M_{n+1}-M_{n}=\alpha \left[\left(n+1\right)^{\beta}-n^{\beta}\right]=\alpha \left[n^{\beta}\left(1+\frac{1}{n}\right)^{\beta}-n^{\beta}\right],
\nonumber
\\
&=\alpha \beta n^{\beta -1}.
\end{align}
Hence the quasi-normal mode frequency is given by,
\begin{align}
\bar{\omega}=\frac{\alpha \beta}{\hbar} n^{\beta -1}=\frac{\alpha \beta}{\hbar} \left(\frac{M_{n}}{\alpha }\right)^{\frac{\beta -1}{\beta}}.
\end{align}
The area turns out to be,
\begin{align}
A=4\pi (2M)^{2}=16\pi M^{2}=16\pi \alpha ^{2}n^{2\beta}.
\end{align}
Thus, entropy and hence the multiplicity of level can be given by,
\begin{align}
S=4\pi \alpha ^{2}n^{2\beta};\qquad g(n)=\exp (S)=\exp \left(4\pi \alpha ^{2}n^{2\beta}\right).
\end{align}
Now we will choose $\alpha$ to be,
\begin{align}
\alpha ^{2}=\frac{1}{4\pi }\ln K; \qquad g(n)=K^{n^{2\beta}}.
\end{align}
Hence $2\beta$ has to be integer which leads to,
\begin{align}
\beta =\frac{1}{2},1, \frac{3}{2},2,\ldots
\end{align}
Therefore, quasi-normal modes turn out to have the following frequencies,
\begin{align}
\bar{\omega}\sim M_{n}^{-1}(\beta =\frac{1}{2});\qquad \bar{\omega}\sim \textrm{constant}(\beta =1),\qquad \bar{\omega}=M_{n}^{1/3}(\beta =\frac{3}{2}).\ldots
\end{align}
Thus, we see that that in this scheme the smallest frequency emitted by the hole scales inversely as its mass. All subsequent frequencies have larger values so the minimum separation remains of the order of the thermal maxima making the spectra discrete. We shall see below that such a feature is borne out by other geometric quantities like area and volume.
\subsection{Area Quantization} 

Under this scheme of quantization, the area becomes one of the fundamental variable which becomes quantized. such a quantization appears readily in the literature on quantum black holes\footnote{In particular, in LQG at the kinematic level, quantization of area element of a spatial slice in spacetime in terms of {\it  spin quantum numbers} is known to indicate ultraviolet finiteness of the theory. We show that even this scheme of quantization reveals a sparse emission profile for black hole.} \cite{Rovelli:2004tv,Bojowald:2010qpa,Perez:2004hj}. 

Again we consider a simple enough quantization of the area of black hole horizon. Most general quantization results can be argued again on the lines discussed in section 2 (see \ref{Arb_Spec} in particular). Therefore, we consider that area of the black hole is quantized such that,
\begin{align}
A_{n}=\alpha n^{\beta}, \label{A-Qunatum}
\end{align}
where $n$ is an integer depicting the quantum level \footnote{In fact, in order to argue for the denseness of Hawking spectra in LQG, it is suggested \cite{Rovelli:2004tv,Bojowald:2010qpa,Perez:2004hj} that the so called spin network punctures the horizon with high spin quantum numbers. In that case the area spectrum approximately follows a quantization like that given in \ref{A-Qunatum} with $\beta=1$. We argue that such an approximation is of a little help as far as densitization of the emission profile is concerned.}. First we consider the black hole as a single highly excited state of such a theory.  A macroscopic description can follow on previous lines. Thus, if the black hole made a transition between $(n+1)$th and nth area level the area change is given by,
\bea
A=16\pi M^{2};\qquad \Delta A=32\pi M \Delta M=32\pi M  \bar{\omega}; \nonumber\\
\Delta A=A_{n+1}-A_{n}=\alpha \beta  n^{\beta -1}.
\eea
This immediately leads to the following expression for quasi normal frequency,
\begin{align}
\bar{\omega}=\frac{\alpha \beta \hbar n^{\beta -1}}{32\pi M \hbar}=\left(\frac{\alpha \beta}{32\pi}\right)\left(\frac{16\pi}{\alpha \hbar}\right)^{\frac{\beta -1}{\beta}}M^{\frac{\beta -2}{\beta}}.
\end{align}
Now, the entropy and multiplicity of nth level can be given by,
\begin{align}
S=\frac{\alpha}{4} n^{\beta};\qquad g(n)=\exp (S)=\exp \left(\frac{\alpha}{4} n^{\beta}\right).
\end{align}
This requires $\alpha =4\ln K$ and $\beta$ to be integer in order to make $g(n)$ integral. Thus quasi-normal modes have the following scaling relations,
\begin{align}
\bar{\omega}\sim M^{-1}(\beta =1), ~~\sim \textrm{constant}(\beta =2), ~~\sim M^{1/3} (\beta =3).\ldots
\end{align}
Thus, clearly such a scheme does not make the spectrum dense. As before the minimum frequency is bounded from below by $\bar{\omega} \propto 1/M$.
\subsection{Volume Quantization}

Lastly, on a 3-dimensional surface let us consider the model where the volume form gets quantized. Classically, the volume of interior of black hole is a foliation dependent quantity \cite{DiNunno:2008id}. However for each of the foliations, the volume remains proportional to $M^3$. In some co-ordinates, the interior becomes dynamic so  the interior geometry evolves in time. However, at each constant time hyper-surface the volume remains proportional to $M^3$. The constant of proportionality does not play much of a role, so we make a simple choice.
Also, as before we only demonstrate the case for a highly excited state such that the above mentioned classical notions are applicable approximately.  Thus, the volume of the $n-$th level can be quantized as,
\begin{align}
V_{n}=\alpha n^{\beta}=\frac{32\pi}{3}M^{3}.
\end{align}
If the black hole makes a transition between $(n+1)$th and $n$th level then the associated volume change is given by, 
\begin{align}
\Delta V=\alpha \beta n^{\beta-1}=32\pi M^{2}\Delta M.
\end{align}
Again the frequency of emission is obtained by the relation $\Delta M=\hbar \bar{\omega}$, the quasi normal mode frequency turns out to be,
\begin{align}
\bar{\omega}=\frac{\alpha \beta}{32\pi \hbar}\left(\frac{32\pi}{3\alpha}\right)^{\frac{\beta -1}{\beta}}M^{\frac{\beta -3}{\beta}}.
\end{align}
The area and hence the entropy leading to the multiplicity turns out to be,
\begin{align}
A=16\pi \left(\frac{3\alpha}{32\pi}\right)^{\frac{2}{3}}n^{\frac{2\beta}{3}};\qquad g(n)=\exp (S)=\exp \left(4\pi \left(\frac{3\alpha}{32\pi}\right)^{\frac{2}{3}}n^{\frac{2\beta}{3}}\right).
\end{align}
The multiplicity would be integer only if $2\beta /3$ is integral, i.e., $\beta =3/2,3,9/2,\ldots$ and $\alpha$ is chosen appropriately. Then the quasi-normal modes behave as,
\begin{align}
\bar{\omega}\sim M^{-1}(\beta =3/2), ~~\sim \textrm{constant} (\beta =3), ~~\sim M^{1/3}(\beta =9/2),\ldots
\end{align}
Thus, we see that the quasi normal mode frequencies in all the three cases follow identical behavior, i.e., it has the minimal separation of the order of $1/M$. It is a trivial extension of the arguments in section 2,  to see the sparseness persists even if the occupancy across the quantum levels is of the kind $n_j \neq \delta_{ij}$ with arbitrary spectral distribution function.
  
\bibliography{Gravity_1_full,Gravity_2_partial,Brane}

\bibliographystyle{./utphys1}
\end{document}